# Review of borophene and its potential applications

Zhi-Qiang Wang,[1,2] Tie-Yu Lü,[1] Hui-Qiong Wang,[1,3] Yuan Ping Feng[2,*], Jin-Cheng Zheng,[1,3,4,*]

[1] Department of Physics, and Collaborative Innovation Center for Optoelectronic Semiconductors and Efficient Devices, Xiamen University, Xiamen 361005, China

[2] Department of Physics, National University of Singapore, Singapore 117542, Singapore

[3] Institute of Artificial Intelligence, Xiamen University Malaysia, 439000 Sepang, Selangor, Malaysia.

[4] Fujian Provincial Key Laboratory of Theoretical and Computational Chemistry, Xiamen University, Xiamen 361005, China

*E-mail: phyfyp@nus.edu.sg, jczheng@xmu.edu.cn

## Abstract

Since two-dimensional boron sheet (borophene) synthesized on Ag substrates in 2015, research on borophene has grown fast in the fields of condensed matter physics, chemistry, material science, and nanotechnology. Due to the unique physical and chemical properties, borophene has various potential applications. In this review, we summarize the progress on borophene with a particular emphasis on the recent advances. First, we introduce the phases of borophene by experimental synthesis and theoretical predictions. Then, the physical and chemical properties, such as mechanical, thermal, electronic, optical and superconducting properties are summarized. We also discuss in detail the utilization of the borophene for wide ranges of potential application among the alkali metal ion batteries, Li-S batteries, hydrogen storage, supercapacitor, sensor and catalytic in hydrogen evolution, oxygen reduction, oxygen evolution, and $CO_2$ electroreduction reaction. Finally, the challenges and outlooks in this promising field are featured on the basis of its current development.

# 1. Introduction

With the boom of graphene[1-11], two-dimensional (2D) materials, such as graphene, BN[12], silicene[13-15], germanene[16], phosphorene[17], transition metal dichalcogenides[18-23], arsenene[24], and antimonene[24, 25] have attracted dramatically increasing interest in the past decade. Plenty of 2D materials have been synthesized or theoretically predicted. Due to the unique physical and chemical properties (such as linear band structure near the Fermi level, high electrical, thermal conductance and its stiffness), 2D materials show vast application prospect in electronic devices, energy storage and utilization.

Borophene, a single layer of boron atom, has been synthesized recently on a silver substrate under ultrahigh-vacuum [26-28]. All four phases (2-*Pmmn*, $\beta_{12}$, $\chi_3$ and honeycomb phases as shown in Figure 1) of borophenes that have been synthesized in the experiments are metallic. Since then, numerous experimental and theoretical studies of the mechanical properties, electronic structure, lattice thermal conductivity, superconducting properties, optical properties, atomic adsorption, and surface reactivity of borophene have been reported [29-50]. Borophene shows some unique physical and chemical properties. For instance, the 2-*Pmmn* phase of borophene possesses a buckled structure with the adjacent row boron atoms corrugating along the zigzag direction. Along the other in-plane direction (armchair direction), the atomic structure is un-corrugated. Interestingly, the Poisson's ratios along both in-plane directions are negative. Highly anisotropic mechanical properties have been observed. The Young's modulus along the armchair direction can be up to 398 N/m, which is even larger than that of graphene. However, the Young's modulus along the zigzag direction is only 170 N/m [26]. Due to the highly anisotropic crystal structure, the electronic band structure also shows high anisotropy. Along the armchair direction, the band structure shows metallic character; however, along the zigzag direction, a large band gap is observed. Ultrafast surface ion transport along the armchair direction has been found. The energy barrier of Li/Na/K/Mg/Al ion transport on the 2-*Pmmn* phase of borophene is about 1.9-39.24 meV. Furthermore, high capacity and excellent electronic conductivity are confirmed. Combined with the high capacity, outstanding electronic and ionic conductivity, borophene shows vast application prospect in metal ion batteries as an anode material. Furthermore, it has been reported that borophene can be a catalyst in hydrogen evolution, oxygen reduction, oxygen evolution and $CO_2$ electroreduction reactions with high catalytic performance. The hydrogen molecule adsorption on the metal-decorated borophene has been studied. The results show that the hydrogen storage capacity is impressive due to the light mass of the boron element and metal atom decoration.

In this review, we first introduce the experimental synthesis and theoretical prediction of borophene. Then, we focus on the physical and chemical properties of borophenes, mainly including the mechanical, thermal, electronic, optical and superconducting properties. Finally, we summarize the application in energy storage and utilization, such as the metal ion batteries, hydrogen storage, sensor and catalytic in hydrogen evolution, oxygen reduction, oxygen evolution, and $CO_2$ electroreduction reactions.

## 2. Experimental synthesis and theoretical prediction

Due to the large energy difference between borophene and bulk boron, borophene is difficult to be synthesized. The energy difference between 2-*Pmmn* phase of borophene and the bulk α-rhombohedral boron is about 555 meV/atom[51]. To the best of our knowledge, four phases (2-*Pmmn*, $\beta_{12}$, $\chi_3$ and graphene-like phases) of borophene have been synthesized on Ag or Al (111) substrates under ultrahigh-vacuum conditions [26-28]. Before the experimental synthesis, all the four structures have been predicted by theoretical calculations, showing the power of theoretical calculations. The scanning tunneling microscope (STM) topography images and structure models are shown in Figure 1. For the 2-*Pmmn* phase of borophene, highly anisotropic electronic and mechanical properties have been reported. 2-*Pmmn* phase of borophene is a corrugated configuration with a buckling height of 0.91 Å. Two borophenes, $\beta_{12}$ and $\chi_3$, have been synthesized on Ag (111) substrate. These two borophenes are planar without vertical undulations; however, different periodic boron vacancy distribution patterns have been observed by experiments and first principles calculations. Recently, a graphene-like honeycomb borophene has been synthesized on Al (111) substrate. It is very interesting that graphene-like borophene is more energetically stable on Al (111) surface than that on Ag (111) surface. Compared with graphene, each boron atom is one electron deficient. The charge transfer from the Al (111) substrate to graphene-like borophene is nearly one electron. However, the charge transfer from the Ag (111) substrate to the graphene-like borophene is negligible. Hence, for graphene-like borophene, it is more energetically stable on Al (111) surface that on Ag (111) surface. Successful experimental synthesis of the four borophenes boosts the corresponding fundamental research and enlightens its practical application in nanodevices and energy field.

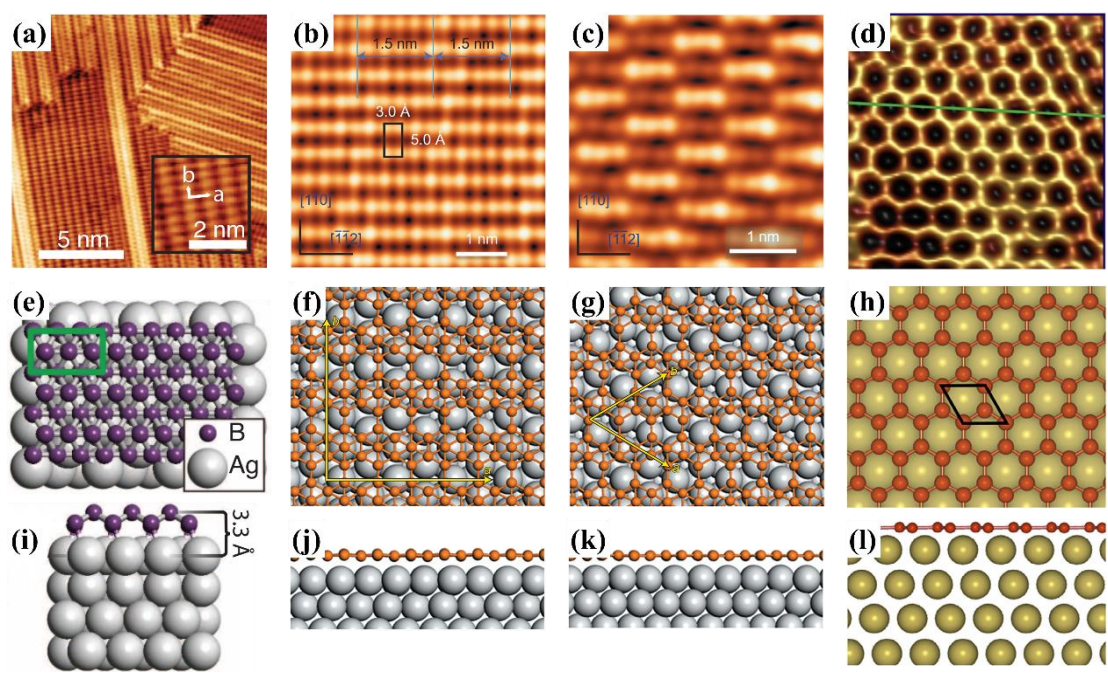

Figure 1. STM topography images (a to d), top view (e to h) and side view (i to l)

structure models of 2-*Pmmn*, $\beta_{12}$, $\chi_3$ and graphene-like borophene on Ag/Al (111) substrate.[26-28]

Plenty of 2D boron allotropes have been reported by theoretical calculations [44, 52-60]. The results show that the stability of borophene is closely related to the hexagon boron vacancy concentration $x$, substrate mediation effect, and chemical modification. Firstly we discuss the effect of the hexagon boron vacancy concentration $x$. Boron vacancy concentration is defined as the ratio between the number of the hexagon boron vacancies and the number of atoms in the original triangular sheet. As shown in Figure 2, the formation energy $E_f$ curve is a V-like function of boron vacancy concentration $x$, which agrees well with other results[61]. At $x=1/9$, the corresponding structure is the most stable one. Intriguingly, the formation energy at two slightly higher vacancy concentration, $x=1/8$ and $2/15$, is very close to that at $x=1/9$. A considerable number of boron allotropes are found whose formation energies are within a few meV/atom of the ground-state line. This polymorphism of 2D boron is completely different from other 2D materials: 2D carbon (graphene), Si (silicene), Ge (germanene), boron nitride (h-BN) and black phosphorus (phosphorene). Graphene, h-BN, silicene, and germanene display a distinct honeycomb structure.

Besides the boron vacancy concentration, the metal substrate mediation plays a key role in the stability of borophenes. Yakobson *et al*. proposed that the ground state structure of borophene in freestanding state and on a substrate is not invariable. As shown in Figure 3, on the Ag (111) surface, $\beta_{12}$ phase of borophene ($x=1/6$) is most stable. The second and third stable structures appear at $x=1/8$ and $1/12$. The corresponding energy difference with respect to the ground state ($\beta_{12}$-borophene) of the second and third stable structures are only 0.4 and 2.1 meV[62], respectively, indicating that borophenes on Ag (111) surface remain polymorphic. However, on the Cu (111) surface, all three sheets with the lowest formation energy $E_f$ appear at $x=1/6$. The corresponding energy differences with respect to the ground state of the second and third stable structures are 15.1 and 28.9 meV[62], respectively, which are much larger than that on Ag (111) surface. This finding shows that it is more likely to form specific borophene structure on Cu (111) surface. In other words, the degree of structural polymorphism of borophene decreased.

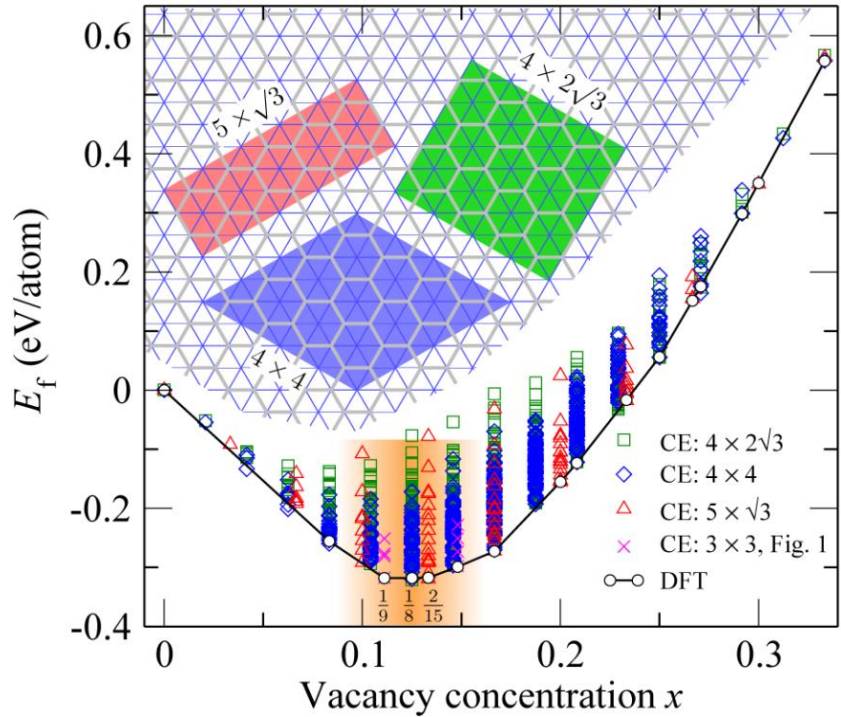

Figure 2. Formation energies $E_f$ of borophenes as a function of boron vacancy concentration $x$. Reproduced from Ref. [52]

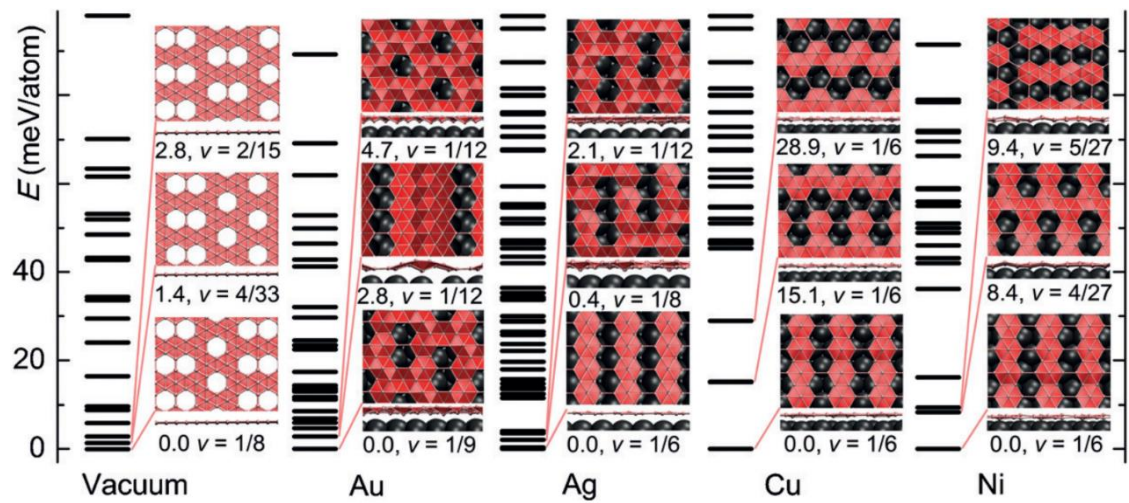

Figure 3. Configurational energy spectra of freestanding and Au, Ag, Cu, and Ni substrates supported borophenes. Reproduced from Ref. [62]

In addition to the boron vacancy concentration and the effect of metal substrate,

functionalization is an effective approach to tune the physical and chemical properties of 2D materials. For example, hydrogenation can open the band gap of graphene, silicene, and germanene. More specifically, before the hydrogenation, graphene, silicene, and germanene are zero gap semiconductors. However, after the hydrogenation, the band gaps of chair-like configuration fully hydrogenated graphene, silicene and germanene are 4.6, 4.0 and 3.6 eV, respectively [63-65]. On the other hand, hydrogenation also has a strong effect on the crystal structure of graphene. Graphene is a planar structure. Fully hydrogenated graphene is a corrugated structure. Hydrogenation can improve the stability of 2D materials. 2-*Pmmn* phase of borophene is not dynamically stable due to the imaginary frequency around the Γ point in the Brillouin zone. After hydrogenation, the imaginary frequency is removed, indicating that hydrogenation can improve the dynamic stability of borophene [66-68]. The electronic band structure also changes due to the hydrogenation. 2-*Pmmn* shows a metallic band structure with some electrons occupy the antibonding states. During the hydrogenation process, a 0.75e charge transfer from boron to hydrogen atoms has been confirmed [67], leading to the results that the in-plane bonding states are completely occupied, antibonding states are empty and the out-plane bonding states are also fully occupied. Finally, fully hydrogenated 2-*Pmmn* phase of borophene possesses a Dirac cone along the X-Γ direction in the Brillouin zone with the Dirac point located at the Fermi level perfectly. Ultrahigh Fermi velocity of $3.5×10^6$ m/s[67], which is even larger than that of graphene, has been observed. The Young's modulus of the fully hydrogenated 2-*Pmmn* phase of borophene along the two in-plane directions are 172.24 and 110.59 N/m, respectively [68]. The ultimate tensile strains along the zigzag and biaxial directions can be up to 0.30 and 0.25, respectively [68]. Fermi velocities can be tuned in a large range by mechanical strains. Three other hydrogenated borophenes (*C2/m*, *Pbcm*, and *Cmmm* phase of borophane) are proposed by Jiao *et al*. The related atomic structure, electron localization function (ELF) and band structure are shown in Figure 4 and Figure 5. The dynamic stability of the three hydrogenated borophenes has been confirmed by phonon dispersions calculations. No imaginary modes can be found in the phonon dispersions. The C2/m and *Pbcm* phase of hydrogenated borophenes possess Dirac cones with massless Dirac fermions and the Fermi velocities for the *Pbcm* and *Cmmm* structures are even higher than that of graphene[66]. More intriguingly, the *Cmmm* borophane shows a Dirac ring (as shown in Figure 5(f)), which is the first report among the boron-based 2D materials.

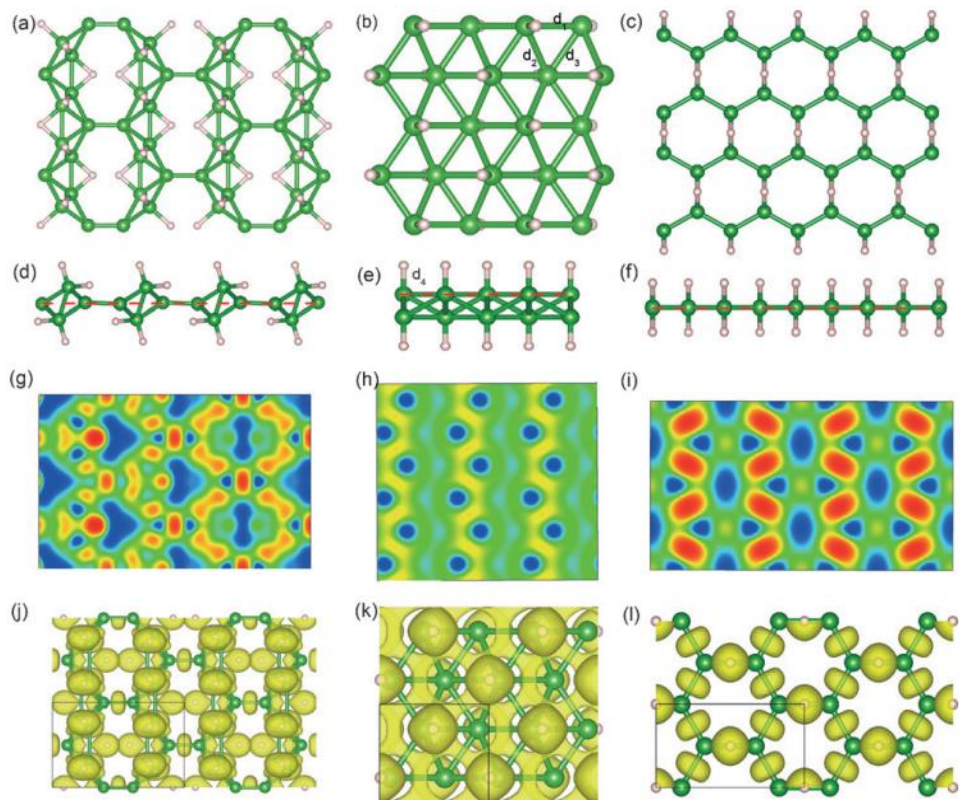

Figure 4. (a)-(f) Top and side view atomic structure, (g)-(i) 2D ELF figure along (001) direction, (j)-(l) 3D ELF (isovalue=0.75) of C2/m, Pbcm, and Cmmm borophanes. Reproduced from Ref. [66]

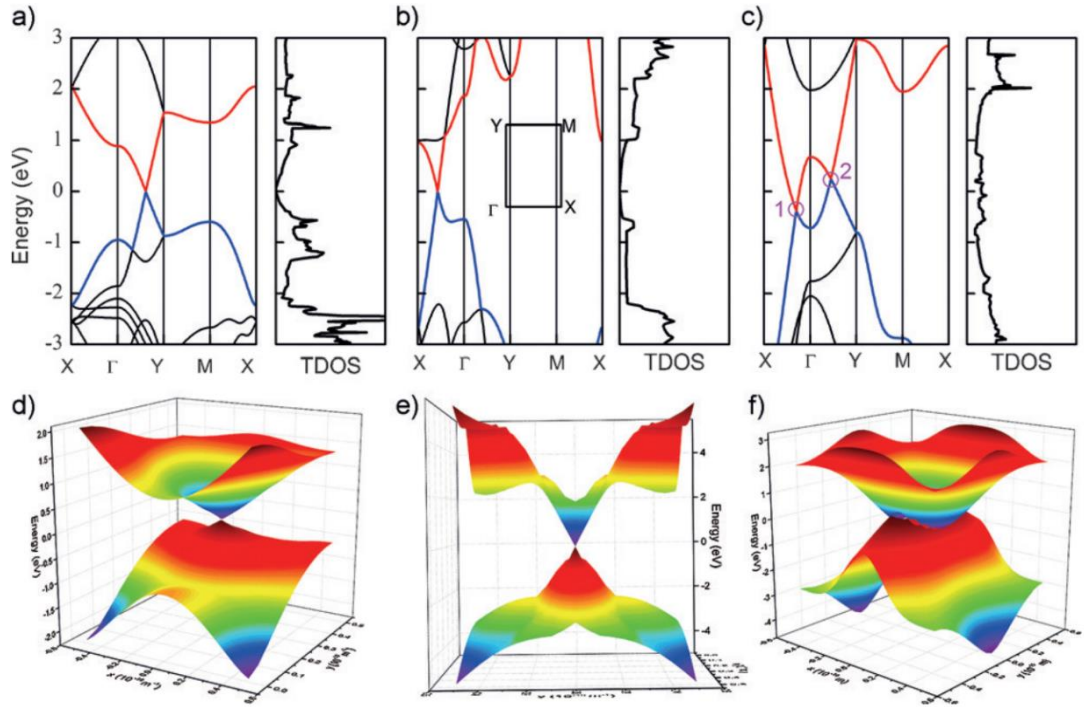

Figure 5. (a)-(c) Band structures and density of states and (d)–(f) the 3D band plots for C2/m, Pbcm, and Cmmm borophanes. Reproduced from Ref. [66]

# 3 Physical and chemical properties

## 3.1 Mechanical properties

Tremendous interest is focused on the mechanical properties of borophenes [45, 53, 69-72]. Due to the strong B-B bonds and unique atomic structure, ultrahigh mechanical modulus of borophene has been reported. The elastic constants $c_{ij}$, layer modulus $\gamma$, shear modulus G, Young's modulus Y in N/m, and Poisson's ratio $\nu$ of borophene, borophane and common 2D materials are listed in Table 1. The Young's modulus of *Pmmm* phase of borophene along the armchair direction can be up to 574.61 N/m, which is much larger than that of graphene (338.08 N/m) [68]. 2-*Pmmn* phase of borophene shows highly anisotropic mechanical properties and negative Poisson's ratio. The Young's modulus along the armchair direction is 398 N/m[34], also larger than that of graphene. The Young's modulus along the zigzag direction (170 N/m) is much smaller. Due to the corrugated structure, the in-plane Poisson's ratios (-0.04 along the armchair direction and -0.02 along the zigzag direction) are negative. The critical strains of borophene and other related 2D materials are listed in Table 2. The ultimate strains of 2-*Pmmn* phase of borophene along the zigzag and armchair directions are 0.15 and 0.10, respectively. The corresponding stresses are 12.95 and 27.46 N/m, respectively [73]. When the strain along the zigzag direction is 0.1, the corresponding stress is only about 11.1 N/m, which is much lower than that under the same strain value along the armchair direction. Due to the compensation of the dihedral angle and B-B bond angles along the zigzag direction, it is easier to apply strain along the zigzag direction. Along the biaxial direction, the ultimate strain is 0.13, and the corresponding stress is about 23 N/m.

The mechanical properties of borophene are affected by the borophene vacancy defect, chemical modification, the number of layers and temperature. Each of these aspects is discussed in the following.

Firstly, the effects of the boron vacancy defect on the mechanical properties were first reported by Mortazavi et al. [70]. Their results show that the Young's modulus along the armchair direction decreases almost linearly with increasing defect concentration. For $\beta_{12}$ and $\chi_3$, the Young's modulus along the armchair direction is 179.00 and 198.50 N/m, respectively, which are much smaller than that of the perfect 2-*Pmmn* phase of borophene (398 N/m). However, along the zigzag direction, the effect of boron vacancy concentration on the Young's modulus is relatively small. Interestingly, in contrast to the negative Poisson's ratio of the defect-free 2-Pmmn phase of borophene, the Poisson's ratios of the α sheet, $\beta_{12}$ and $\chi_3$ phases are positive. By analyzing the crystal structure, we can attribute the negative Poisson's ratio of 2-*Pmmn* phase of borophene to its corrugated structure. However, all the critical strains for α sheet, $\beta_{12}$ and $\chi_3$ along the two in-plane directions are larger than that of the defect-free 2-*Pmmn* phase of borophene.

Chemical modification has a strong effect on the mechanical properties of borophene. The Young's modulus of the fully hydrogenated 2-*Pmmn* phase of borophene along the two in-plane directions are 172.24 and 110.59 N/m, respectively [68]. The dramatical decrease can be understood from the atomic structures. Before hydrogenation, the B-B bond length along the armchair direction is 1.616 Å, which is stretched to 1.941 Å up on hydrogenation. This bond elongation leads to the decrease in the bond strength, and a reduction in the Young's modulus. On the other hand, hydrogenation increases the critical strain. The critical strains of 2-*Pmmn* phase of borophene along the *a*-, *b*- and biaxial directions are 0.08, 0.15 and 0.08, respectively. After the hydrogenation, the critical strains increase to 0.12, 0.30 and 0.25, respectively [68]. Fluorination has a similar effect on the mechanical modulus of 2-*Pmmn* phase of borophene. For $B_4F$, the Young's modulus along the two in-plane directions are 274.13 and 183.04 N/m, respectively. For $B_2F$, the corresponding values are 129.93 and 200.46 N/m, respectively[74], showing a dramatic decrease of the Young's modulus along the armchair direction.

The number of layers in multi-layer 2-*Pmmn* phase of borophene also has a strong effect on its mechanical properties. From one layer to four layers, the Young's modulus of 2-*Pmmn* phase of borophene along the armchair direction is 397, 380, 360 and 338 N/m, respectively. Along the zigzag direction, the corresponding Young's modulus is 158, 144, 141 and 136 N/m, respectively. The Young's moduli along the two in-plane directions decrease with increasing number of layers. A similar phenomenon has been observed in black phosphorene[75].

The Young's modulus of borophene was found to decrease with temperature [72, 76]. For the borophene with a boron concentration *x* of 4/27, the Young's modulus along the zigzag direction is 180 N/m at 1 K, which is reduced to 87 N/m at 600 K.

Due to the structural polymorphism of borophene, unique mechanical properties (negative Poisson's ratio, ultrahigh Young's modulus, high anisotropy) have been found. Furthermore, in our previous work, extraordinary mechanical properties have been found in boron nanoribbon networks. Due to the breaking of the B–B bonds and the rotation of the local structures, structural phase transition occurs under the *a*-, *b*- direction uniaxial and biaxial strains. Subsequently, successive phase transition generates ultrahigh critical tensile strains. The critical tensile strains of χ-h1 boron nanoribbon network along the *a*- and *b*-direction are 0.51 and 0.41, respectively. Along the biaxial direction, the critical strain is 0.84, which is ultrahigh compared to other 2D materials. Furthermore, the structural phase transition can effectively decrease the strain energy, making the boron nanoribbon network easier to stretch. Surprisingly, outstanding electronic transport properties of χ-h1 boron nanoribbon network has been verified by Yi *el al.* [77] The outstanding flexibility and excellent electronic conductivity lead to a vast promising application of borophene in flexible electronic devices.

Table 1. Elastic constants $c_{ij}$, layer modulus γ, shear modulus G, Young's modulus Y in N/m, and Poisson's ratio ν of borophene, borophane and common 2D materials, respectively.

| | $c_{11}$ | $c_{22}$ | $c_{12}$ | $c_{66}=G$ | $Y_a$ | $Y_b$ | $v_a$ | $v_b$ |
|---|---|---|---|---|---|---|---|---|
| ML 2-*Pmmn*[68] | 379.00 | 162.50 | -2.00 | 87.00 | 378.97 | 162.49 | **-0.012** | **-0.005** |
| [34] | 398.00 | 170.00 | -7.00 | 94.00 | 398.00 | 170.00 | **-0.040** | **-0.020** |
| [78] | 396.60 | 158.40 | -3.47 | 86.50 | 397 | 158 | **-0.022** | **-0.009** |
| BL 2-*Pmmn*[78] | 380.0 | 143.8 | 7.44 | 75.2 | 380 | 144 | 0.052 | 0.020 |
| 2-*Pmmn* 3Layer [78] | 361.1 | 141.5 | 11.29 | 73.5 | 360 | 141 | 0.080 | 0.031 |
| 2-*Pmmn* 4Layer [78] | 337.9 | 136.0 | 12.37 | 70.8 | 338 | 136 | 0.091 | 0.037 |
| B₄F[74] | 275.06 | 183.66 | 13.07 | 79.63 | 274.13 | 183.04 | 0.07 | 0.05 |
| B₂F[74] | 130.12 | 200.76 | 6.17 | 57.35 | 129.93 | 200.46 | 0.03 | 0.05 |
| α sheet[68] | 219.00 | 219.00 | 43.00 | 88.00 | 210.56 | 210.56 | 0.196 | 0.196 |
| β₁₂[68] | 185.50 | 210.50 | 37.00 | 68.50 | 179.00 | 203.12 | 0.176 | 0.199 |
| χ₃[68] | 201.00 | 185.00 | 21.50 | 60.50 | 198.50 | 182.70 | 0.116 | 0.107 |
| *Pmmm*[68] | 333.50 | 576.00 | 21.50 | 157.00 | 332.70 | **574.61** | 0.037 | 0.064 |
| 8-*pmmn*[68] | 249.00 | 322.50 | 15.50 | 108.00 | 248.26 | 321.54 | 0.048 | 0.062 |
| BNRN χ-h1[79] | 138.25 | 138.25 | 39.25 | 50.50 | 127.12 | 127.12 | 0.284 | 0.284 |
| C-borophane[68] | 175.77 | 112.86 | 19.97 | 28.46 | 172.24 | 110.59 | 0.177 | 0.144 |
| B-borophane[80] | 197.00 | 83.00 | 22.00 | 53.50 | 191.17 | 80.54 | 0.265 | 0.112 |
| TCB-borophane[80] | 160.00 | 122.00 | 29.00 | 70.00 | 153.11 | 116.74 | 0.238 | 0.181 |
| T-borophane[80] | 129.50 | 154.00 | 27.50 | 79.50 | 124.59 | 148.16 | 0.179 | 0.212 |
| W-borophane[80] | 179.50 | 159.00 | 21.00 | 84.00 | 176.73 | 156.54 | 0.132 | 0.117 |
| Graphene[68] | 348.75 | 348.75 | 61.00 | 143.88 | 338.08 | 338.08 | 0.175 | 0.175 |
| Ref. [81] | 352.70 | 352.70 | 60.90 | 145.90 | 342.18 | 342.18 | 0.173 | 0.173 |
| Ref. [82] | 358.10 | 358.10 | 60.40 | 148.90 | 347.91 | 347.91 | 0.169 | 0.169 |
| BN[81] | 289.8 | 289.8 | 63.7 | 113.1 | 275.8 | 275.8 | 0.22 | 0.22 |
| Phosphorene[79] | 186.80 | 44.61 | 31.23 | 41.82 | 167.88 | 39.39 | 0.70 | 0.167 |
| Silicene[81] | 68.3 | 68.3 | 23.3 | 22.5 | 60.6 | 60.6 | 0.341 | 0.341 |
| Germanene[81] | 46.4 | 46.1 | 13.1 | 16.7 | 42.7 | 42.7 | 0.282 | 0.282 |

Table 2. Critical strain of borophene and common 2D materials. For 8-*Pmmn* phase of borophene, we define the direction along the atomic wrinkle direction as the zigzag direction and the corresponding perpendicular direction as the armchair direction.

| | Armchair /$a$-direction | Zigzag /$b$-direction | Biaxial |
|---|---|---|---|
| ML 2-*Pmmn* [83] | 0.08 | 0.15 | 0.08 |
| Ref.[73] | 0.10 | 0.15 | 0.13 |
| BL 2-*Pmmn*[78] | 0.14 | 0.16 | 0.13 |
| 2-*Pmmn* 3Layer [78] | 0.14 | 0.16 | 0.15 |
| 2-*Pmmn* 4Layer [78] | 0.14 | 0.15 | 0.14 |
| $\beta_{12}$[83] | 0.20 | 0.12 | |
| Ref.[73] | 0.21 | 0.18 | 0.16 |
| $\chi_3$[83] | 0.11 | 0.21 | |
| 8-*Pmmn*[83] | 0.16 | 0.17 | 0.13 |
| BNRN $\chi$-h1[79] | 0.51 | 0.41 | 0.84 |
| Graphene[84] | 0.19 | 0.24 | 0.24 |
| BN[84] | 0.18 | 0.26 | 0.24 |
| $MoS_2$[83] | 0.26 | 0.18 | 0.22 |
| Phosphorene[83] | 0.27 | 0.33 | |
| Silicene[83] | 0.17 | 0.14 | 0.16 |
| Germanene[85] | 0.20 | 0.21 | |
| Stanene[85] | 0.17 | 0.18 | |

## 3.2 Thermal conductivity

Thermal conductivity is an important physical parameter for the performance and longevity of nano devices. The thermal conductivity and phonon transmission of borophene have been studied [86-92]. Due to the highly anisotropic atomic structure of 2-*Pmmn* phase of borophene, the thermal conductivity is anisotropic. At room temperature, the thermal conductivity of 2-*Pmmn* phase of borophene is ~75.9 and ~147 W/mK along the zigzag and armchair directions, respectively. The effective phonon mean free path is 16.7 and 21.4 nm along the two directions, respectively. For α-sheet borophene, an isotropic material, the thermal conductivity is 14.34 W/mK. Surprisingly, the thermal transport is dominatingly contributed by high-frequency phonon modes. For graphene[93], silicene[93] and phosphorene[94], the thermal transport is dominated by the low-frequency acoustic modes. In addition, a dimensionality cross-over of the phonon transmission of hydrogenated 2-*Pmmn* phase of borophene has been proposed by Li *et al.* [86], i.e., the phonon transmission of low-frequency phonons shows 2D behavior, however, at high-frequency, the phonon transmission shows 1D behavior. The transmission of high-frequency phonons is close to 0 nm$^{-1}$. Due to the excellent mechanical properties and high thermal conductivity, hydrogenated 2-*Pmmn* phase of borophene is promising for the soft thermal channel. The unique dimensionality crossover in phonon transmission provides a good approach to study the effect of phonon population, which is important for thermal transport in low-dimensional systems.

## 3.3 Electronic and optical properties

The electronic structure of 2-*Pmmn* phase of borophene is also highly anisotropic[80]. More specifically, the electronic band structure shows metallic conduction along the Γ-X and Y-S directions (armchair direction). A high group velocity of 6.6×10$^5$ m/s has been confirmed[51]. However, the band structure shows a band gap along the Γ-Y and S-X directions (zigzag direction).

Chemical modification is an effective approach to tune the band structure of 2D materials. Band gap opening has been found in graphene, silicene, and germanene by hydrogenation and fluorination. The band structure of 2-*Pmmn* phase of borophene is metallic. However, for the fully hydrogenated 2-*Pmmn* phase of borophene, the band structure change to a zero-gap semiconductor. A Dirac cone has been found and the Dirac point is perfectly located at the Fermi level. The transition from metallic band structure to zero gap semiconductor band structure has been reported. Moreover, the effect of fluorination on the band structure of 2-*Pmmn* phase of borophene has been studied by first principles calculations. Two stable fluorinated borophene structures, $B_4F$ and $B_2F$, were proposed. For $B_4F$, the metallic electronic band structure is preserved. However, $B_2F$ is an indirect semiconductor with a band gap of 0.4 eV [74]. The thermodynamic stability of $B_4F$ and $B_2F$ at finite temperatures has been evaluated by using first principles molecular dynamics (FPMD) simulation. The results show that both $B_4F$ and $B_2F$ are stable even at 1000 K. The intrinsic semiconducting electronic band structure, the excellent mechanical properties, and the good stability make the fluorinated borophene a promising candidate in semiconductor electronic devices.

Generally speaking, mechanical strain can have a strong effect on the physical and chemical properties of materials[95-98]. The effects of external mechanical strains on the electronic band structure of the *Pbcm* phase of borophane have been investigated. The results show that uniaxial strains along either in-plane direction does not open a gap at the Dirac point. However, a band gap can be open at the Dirac point by a shear strain. At a shear strain of 0.01, the strain-induced band gap was found to be 55 meV[99]. Furthermore, the shear strain induced band gap increases with the increasing shear strain. At a share strain of 0.12, the band gap reaches 538 meV[99]. The excellent dynamical stability of *Pbcm* borophane under the shear stain has been confirmed by the phonon dispersion. Shear strain induced band gap opening was also found in fully hydrogenated 2-*Pmmn* phase of borophene. Similarly, the band gap increases with the increasing shear strain. A similar effect of mechanical strain on the band structure of 8-*Pmmn* phase of borophene has been reported. Here, neither uniaxial nor biaxial strain cannot open a gap at the Dirac point. However, gap opening at the Dirac point can be achieved by a shear strain. Unexpectedly, under the stress-free state, hydrogenated graphene-like borophene possesses a metallic band structure. A metallic to semiconducting band structure transition can be induced by a uniaxial and a biaxial strain. More specifically, under a *b*-direction uniaxial or biaxial strain of 0.2, the band gap is 380 and 120 meV, respectively. The unique effect of mechanical strains on the band structure of hydrogenated graphene-like borophene has been observed. The strain-induced band structure transition provides a new enlightenment for us to tune the band structure of 2D materials.

Anisotropic optical properties have been found in borophene materials. For the absorption and reflectivity coefficient, the positions of the main absorption are different along the two in-plane directions. Along the armchair direction, the absorption regions are around 3.65 and 10.36 eV. However, the main absorption peaks are located at around 1.09, 8.29 and 10.31 eV, respectively[100]. Furthermore, in the visible region, the reflectivity of borophene along the armchair and zigzag directions are <30% and >40%, respectively. Both the transmittance and electrical conductivity of 2-*Pmmn* phase of borophene along the armchair direction is higher than that along the zigzag direction. The optical conductivity of 2-*Pmmn* phase of borophene is very small in the visible region. However, fluorinated 2-*Pmmn* phase of borophene ($B_2F$) shows significant optical conductivity in the visible region, suggesting promising applications in optical devices.

## 3.4 Superconducting properties

Superconductivity properties of borophene have been reported in various studies[29, 54, 61, 101-106]. Zhao et al. carried out a systematic study of borophene with a range of boron vacancy concentration[61]. As shown in Figure 6, it is very intriguing that the superconducting transition temperature $T_c$ shows a V-shaped dependence on the hexagon hole density. The superconducting transition temperature $T_c$ was found to decrease gradually with increasing boron vacancy concentration, up to $x = 1/9$, beyond which $T_c$ increases gradually with the boron vacancy concentration. In addition to the superconducting transition temperature $T_c$, V-shaped dependence on boron vacancy concentration was also found for total energy, electronic density of states $N_F$ at the Fermi level, electron-phonon coupling (EPC) parameter $\lambda$. The inflection points of the above four curves are located at the $x=1/9$. This work provides a useful approach to analyze the stability and superconducting properties of 2D materials. A positive correlation between the density of state at the Fermi level and the superconducting property has been observed in borophene and transition metal intercalated bilayer borophene.

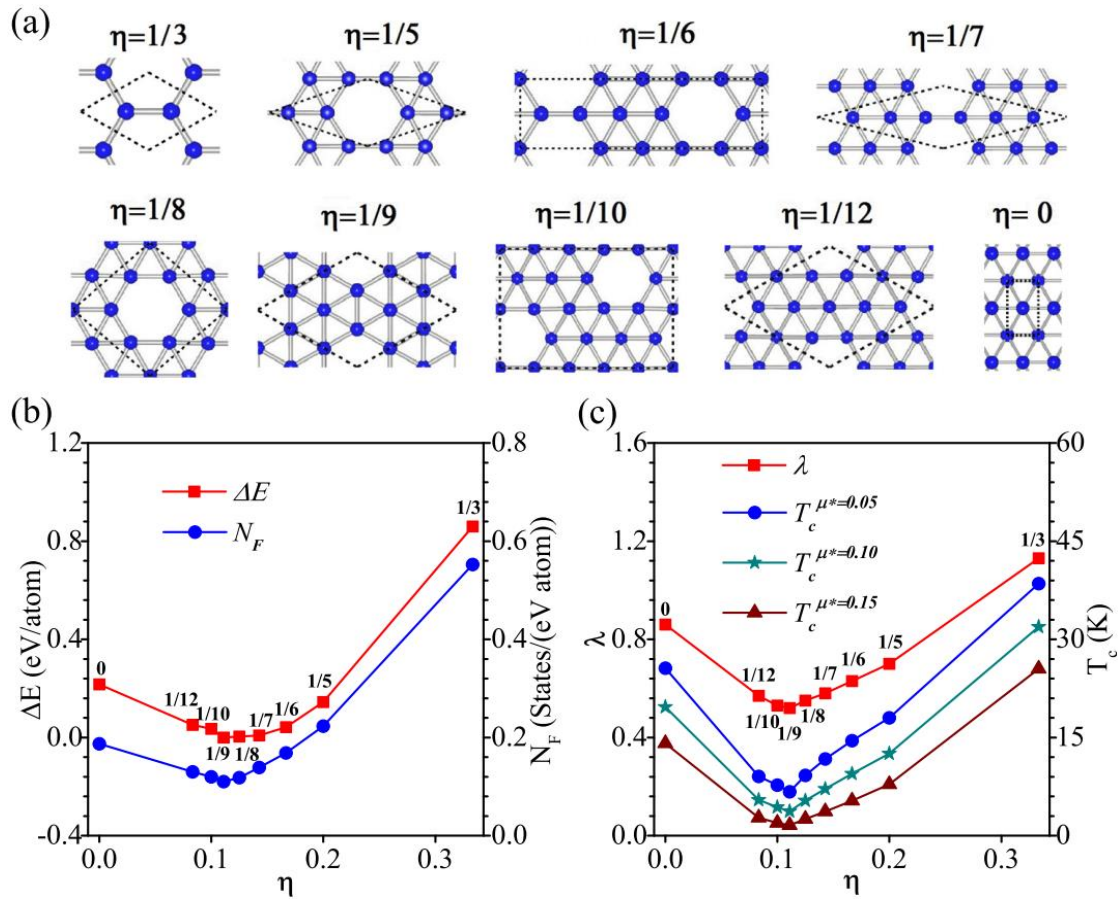

Figure 6. (a) Most stable borophenes under different boron vacancy concentration. (b) Total energy and $N_F$ vs boron vacancy concentration. The total energy of $\alpha$ sheet is set to be zero. (c) $\lambda$ and Tc vs boron vacancy concentration. Reproduced from Ref.[61]

For 2-*Pmmn* phase of borophene, high density of states at the Fermi energy and strong Fermi surface nesting lead to strong electron-phonon coupling. The superconducting transition temperature $T_c$ of freestanding 2-*Pmmn* phase of borophene is about 19 K.

Furthermore, the $T_c$ can be tuned by strain and carrier doping. More specifically, a tensile strain can increase the $T_c$ to 27.4 K, while hole doping can increase the $T_c$ to 34.8 K[35]. On the other hand, a compressive strain or electron doping decreases the $T_c$ of 2-*Pmmn* phase of borophene. The electron-doping induced suppression of the superconductivity properties was also found in the $\beta_{12}$ phase of borophene. The superconducting transition temperature of $\beta_{12}$ phase of borophene is 14 K. Electron doping of 0.1 e per boron atom reduces the $T_c$ to 0.09 K. The impact of strain and carrier doping on the $T_c$ is closely related to the density states of in-plane orbitals near the Fermi surface. The electron-doping induced suppression explains the challenges to experimentally probe superconductivity in substrate-supported borophene.

In addition, the superconductivity properties of Mg intercalated bilayer borophene systems $B_xMgB_x$ (*x*=2~5) have been studied by first principle calculations. Both $B_2MgB_2$ and $B_4MgB_4$ exhibit good phonon-mediated superconductivity with a high $T_c$ of 23.2 and 13.3 K, respectively[101]. In the Mg-B systems, the $T_c$ decreases with decreasing in-plane projected density of states at the Fermi level, rather than with total density of states. This finding provides an effective approach to tune the superconductivity of free-standing, substrate-supported borophene and metal intercalated bilayer borophenes. Besides the above two avenues (strain and carrier doping) that can tune the $T_c$, the superconductivity properties of borophene are also closely related to its intrinsic crystal structure. Borophene shows a certain application prospect in superconductivity with a high superconducting transition temperature.

# 4 Application of borophene as energy materials

## 4.1 Alkali metal ion batteries

For electrode materials, high capacity, suitable voltage, good electronic and ionic conductivity are crucial parameters. Borophene is the lightest 2D material. Furthermore, high surface activity of borophene has been reported. Both of these two factors are beneficial to achieve ultrahigh capacity for Li/Na/K/Mg/Al storage. The metallic band structure is also good for electronic conductivity. Hence, borophene can be a promising electrode material for Li/Na/K/Mg/Al ion batteries.

Zhao *et al*. firstly proposed 2-*Pmmn* borophane as a good anode materials for Li ion batteries. Due to the high surface activity of 2-*Pmmn* phase of borophene, the interaction between Li and borophene is strong. The initial intercalation voltage (at low Li adsorption coverage) is 1.12 eV, and decreases with increasing Li adsorption coverage. The fully lithiated phase of 2-*Pmmn* phase of borophene is $Li_{0.75}B$ which corresponds to a theoretical specific capacity of 1860 mAh/g, much larger than that of graphite (372 mAh/g)[107], silicene (954 mAh/g)[108], phosphorene (433 mAh/g)[109], $VS_2$ (466 mAh/g)[110], $Ti_3C_2$ (320 mAh/g)[111] and $Li_4Ti_5O_{12}$ (175 mAh/g)[112]. Furthermore, as shown in Figure 7, the Li ion migration barrier along the armchair direction is only 2.6 meV, indicating ultrafast Li ion migration along the armchair direction. The Li ion migration barrier of 2-*Pmmn* phase of borophene is much smaller than that of graphite (450-1200 meV)[107], silicene (230 mAh/g)[108], phosphorene (80 mAh/g)[109], and $Li_4Ti_5O_{12}$ (300 mAh/g)[112]. However, along the zigzag direction, the energy barrier of 2-*Pmmn* phase of borophene is about 325 meV. Hence, the surface Li ion migration shows strong anisotropy. During the whole lithiation process, the electronic structures are metallic characteristic, indicating excellent electronic conductivity of borophene. The ultrahigh ionic conductivity and excellent electronic conductivity guarantee an outstanding rate performance of 2-*Pmmn* borophene during the charging and discharging process.

Applications of 2-*Pmmn* phase of borophene in Na, K, Mg, Ca, Al ion batteries have also been reported. Ultrahigh capacity and ultrafast ion diffusion of 2-*Pmmn* phase of borophene have been identified. More specifically, the capacities of 2-*Pmmn* phase of borophene as an anode material are 6611 and 9917 mAh/g for Mg and Al ion batteries, respectively. The energy barriers of Mg and Al ion migration along the armchair direction are only 11.76 and 39.24 meV, respectively.

Furthermore, doping is a common approach to tune the physical and chemical properties of functional materials [113, 114]. The effects of P doping, hydrogenation and substrate have been studied. Hydrogenation can stabilize 2-*Pmmn* phase of borophene by the charge transfer from borophene to hydrogen atoms. Hence, the interactions strength between Li and hydrogenated borophene decreases. As an anode material for Li ion batteries, the capacity of hydrogenated 2-*Pmmn* phase of borophene is 504 mAh/g, which is much lower than that of 2-*Pmmn* phase of borophene (1860 mAh/g).

In addition, the $\beta_{12}$ and $\chi_3$ phases of borophene also show high capacities in Li, Na, and Mg ion batteries. As the anode material, the capacities of $\beta_{12}$ and $\chi_3$ phases of borophene in Li ion batteries reach 1984 and 1240 mAh/g, respectively, which are much higher than those of graphene, silicene, phosphorene, and $Li_4Ti_5O_{12}$. However, due to the periodically arranged boron vacancies and the structural transition from the corrugated structure to the flat structure, the migration energy barriers of Li, Na and Mg on $\beta_{12}$ and $\chi_3$ phases of borophene are larger than that on 2-*Pmmn* phase of borophene. But the energy barrier is still acceptable as anode materials for Li/Na ion batteries. As listed in Table 3, the migration energy barrier of Na ion (330 meV) for $\beta_{12}$ phase of borophene is much smaller than that of Li ion (660 meV). Similarly, for $\chi_3$ phase of borophene, the migration energy barrier of Na ion (340 meV) is much smaller than that of Li ion (600 meV). Therefore, from the point of view of ionic conductivity and rate performance, borophene with boron vacancies can be expected to perform better as an anode material in Na ion batteries than in Li ion batteries.

Thus, borophene is a promising anode material for Li, Na and Mg ion batteries due to high theoretical specific capacities, excellent electronic conductivity and outstanding ion transport properties[107, 115-121].

Table 3. The theoretical specific capacity (C) and diffusion barriers ($E_a$) of borophene and some common 2D materials as anode materials for alkali metal ion batteries.

| Species | M | C mAh/g | $E_a$ meV | Ref | Species | M | C mAh/g | $E_a$ meV | Ref |
|---|---|---|---|---|---|---|---|---|---|
| 2-*Pmmn* | Li | 1860 | 2.6 | [107] | $\beta_{12}$ | Li | 1880 | 690 | [122] |
| | Li | 3306 | 10.53 | [119] | | Na | 1984 | 330 | [116] |
| | Li | 1239 | 7 | [123] | | Na | 1640 | 340 | [122] |
| | Na | 1218 | 1.9 | [117] | | Mg | 2480 | 970 | [122] |
| | Na | 2341 | 2.55 | [119] | $\chi_3$ | Li | 1240 | 600 | [116] |
| | K | 1377 | 7.61 | [119] | | Li | 1240 | 475 | [124] |
| | Mg | 6611 | 11.76 | [119] | | Li | 2040 | | [122] |
| | Al | 9917 | 39.24 | [119] | | Na | 1240 | 340 | [116] |
| 2-*Pmmn*/Ag | Li | | 33 | [123] | | Na | 1480 | | [122] |
| 2-*Pmmn*-P | Li | 1732 | 328 | [125] | | Mg | 2400 | | [122] |
| 2-*Pmmn*-H | Li | 504 | 210 | [126] | Graphite | Li | 372 | 450-1200 | [107] |
| 2-*Pmmn*-H | Na | 504 | 90 | [126] | Phosphorene | Li | 433 | 80 | [109] |
| $\beta_{12}$ | Li | 1984 | 660 | [116] | Silicene | Li | 954 | 230 | [108] |
| | Li | 1983 | 590 | [124] | $Li_4Ti_5O_{12}$ | Li | 175 | 300 | [112] |

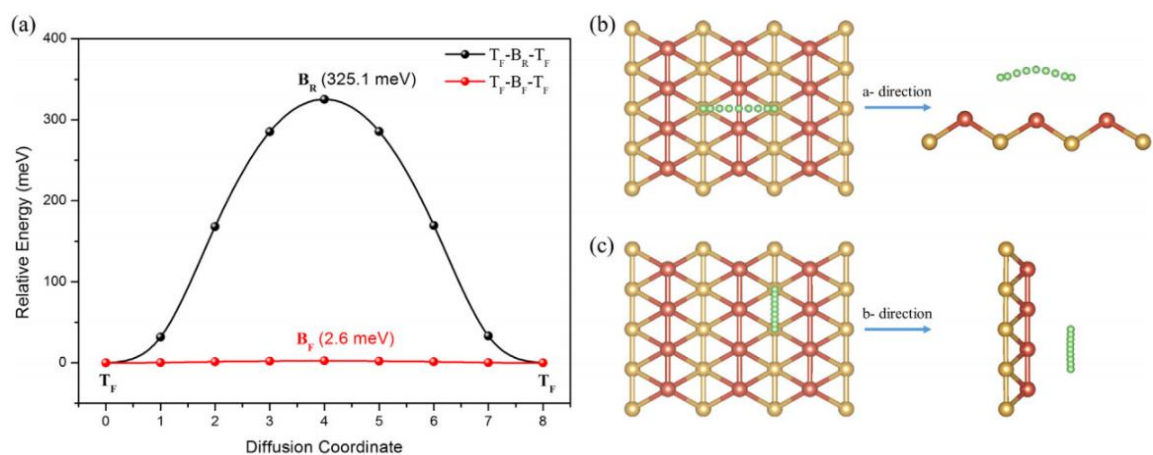

Figure 7.  (a) Energy barriers, and (b), (c) pathways of Li migration along zigzag and armchair direction. Reproduced from Ref.[107]

## 4.2 Li-S batteries

The advance of lithium-sulfur batteries has been hampered by the fast capacity fading induced by the shuttle effect[127]. Suitable sulfur anchoring materials can suppress the shuttle effect and improve the cycling performance. The interaction between the anchoring material and lithium polysulfides should not be too strong or too weak. As listed in Table 4, for graphene, the adsorption energies of $Li_2S_4$, $Li_2S_6$, and $Li_2S_8$ are 0.65, 0.72 and 0.73 eV, respectively. Due to the weak interaction, graphene is not a suitable anchoring material for lithium polysulfides. It is imperative to search for more suitable anchoring materials for lithium batteries.

The application of borophene as the potential anchoring materials for lithium-sulfur batteries has been studied by first principles calculations [128-130]. Zhao *et al.* concluded that the 2-*Pmmn* phase of borophene is not a good anchoring material for the excessively strong interaction between lithium polysulfides and 2–*Pmmn* phase of borophene. The adsorption energies of $Li_2S_4$, $Li_2S_6$, and $Li_2S_8$ on 2–*Pmmn* phase of borophene are 6.45, 4.32 and 6.18 eV, respectively. Such strong interactions will lead to the decomposition of lithium polysulfides, resulting in the irreversible sulfur loss during the charging and discharging processes. Unexpectedly, the interaction between lithium polysulfides and the $\chi_3$ phase of borophene is much smaller than that between the lithium polysulfides and 2-*Pmmn* phase of borophene, but larger than that between lithium polysulfides and graphene. The adsorption energies of $Li_2S_4$, $Li_2S_6$, and $Li_2S_8$ on $\chi_3$ phase of borophene are 2.67, 2.53 and 2.87 eV, respectively, indicating that $\chi_3$ phase of borophene is a suitable anchoring material for lithium-sulfur batteries. The suitable adsorption strength is helpful to suppress the shuttle effect and keep their cyclic structure undecomposed during the charging and discharging processes. Furthermore, borophene shows a metallic electronic structure during the whole battery cycling. Hence, $\chi_3$ phase of borophene is a promising anchoring material for lithium-sulfur batteries. In addition, the $\beta_{12}$ phase of borophene also shows good performance as the anchoring material for Lithium-sulfur batteries. The adsorption energies of $Li_2S$, $Li_2S_2$, $Li_2S_4$, $Li_2S_6$, and $Li_2S_8$ on $\beta_{12}$ phase of borophene are 3.34, 2.89, 1.45, 1.53 and 1.36 eV, respectively. Furthermore, during the whole charging and discharging processes, $\beta_{12}$ phase of borophene shows metallic electronic structure. Thus, both the $\chi_3$ and $\beta_{12}$ phases of borophenes are promising anchoring materials for lithium-sulfur batteries.

Table 4. The adsorption energies (in eV) of $Li_2S$, $Li_2S_2$, $Li_2S_4$, $Li_2S_6$ and $Li_2S_8$ on 2-*Pmmn*, $\chi_3$, $\beta_{12}$ borophene, graphene, and phosphorene.

| Species | $Li_2S$ | $Li_2S_2$ | $Li_2S_4$ | $Li_2S_6$ | $Li_2S_8$ | Ref |
|---------|---------|-----------|-----------|-----------|-----------|-----|
| 2-*Pmmn* | | | 6.45 | 4.32 | 6.18 | [131] |
| $\chi_3$ | | | 2.67 | 2.53 | 2.87 | [131] |
| $\beta_{12}$ | 3.34 | 2.89 | 1.45 | 1.53 | 1.36 | [128] |
| Graphene | | | 0.65 | 0.72 | 0.73 | [132] |
| Phosphorene | 2.51 | 1.91 | 1.27 | 1.00 | 1.12 | [133] |

## 4.3 Hydrogen storage

As the lightest 2D material, borophene is also a promising material for hydrogen storage, and can lead to ultrahigh hydrogen storage capacity. This has been investigated in a number of studies [134-146].

For the $\alpha$-sheet of borophene, the adsorption energy of a single $H_2$ molecule is only 0.047 eV. The interaction of $H_2$ molecule with the $\alpha$-sheet phase of borophene shows similarities to its interaction with graphene (0.025 eV) [134]. The adsorption energy is too small for practical hydrogen storage. Metal atom decoration is an effective and feasible approach to strengthen the interaction. After Li decoration of $\alpha$-sheet phase of borophene, the adsorption energy of $H_2$ molecule reaches 0.35 eV, a significant increase compared to that without the Li decoration. As shown in Figure 8, up to three $H_2$ molecules can be adsorbed on a Li atom. This is equivalent to a hydrogen storage capacity of 10.7 wt %, which is much higher than that of Li-decorated silicene (6.3 wt %) and $MoS_2$ (4.8 wt %). In Table 5, we list the adsorption energies and hydrogen storage capacities of the defective 2-*Pmmn* phase of borophene, AlN, metal atom decorated borophenes, graphene, graphyne, silicene, $BC_7$, and $MoS_2$. High hydrogen storage capacity of 10.75 (12.68) wt% for Li (Ca)-decorated $\alpha$-sheet phase of borophene and 15.26 wt% for Li-decorated borophene ($x$=1/8) have been reported, indicating that alkalis or alkaline metal decorated borophene shows the application prospect in hydrogen storage.

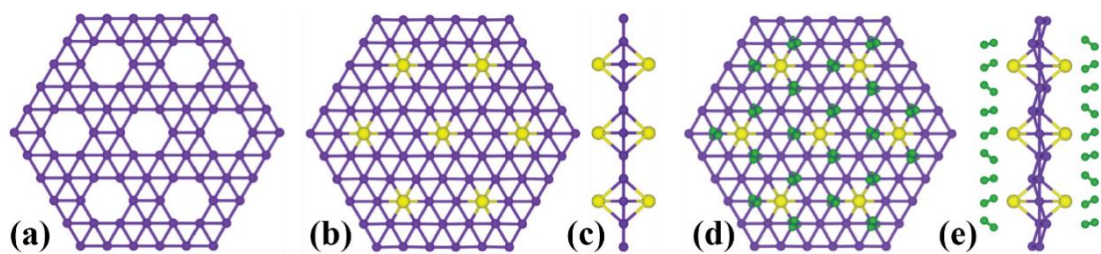

Figure 8. Crystal structures of α sheet borophene without (a) and with (b), (c) Li adsorption. (d) and (e) Crystal structures of Li-decorated α sheet borophene with hydrogen molecules adsorption. Reproduced from Ref.[134]

Table 5. The adsorption energies $E_b$ and hydrogen storage capacity of the metal atom (M) decorated borophenes, defective 2-*Pmmn* borophene, graphene, graphyne, silicene, BC$_7$, MoS$_2$, and AlN. The η value in the bracket is the boron vacancy concentration.

| Species | M | $E_b$ (eV) | Capacity (wt %) | Ref |
|---|---|---|---|---|
| α-sheet | Li | 0.15 | 10.75 | [134] |
| | Na | 0.07 | 8.36 | [134] |
| | K | 0.06 | 2.39 | [134] |
| | Ca | 0.19 | 12.68 | [138] |
| Borophene (η=1/8) | Li | 0.23 | 15.26 | [144] |
| Borophene (η=1/7) | Li | 0.35 | 9.22 | [143] |
| β$_{12}$ | Li | 0.22 | 10.85 | [147] |
| | Ca | 0.24 | 9.5 | [137] |
| 2-*Pmmn* | Li | ~0.11 | 6.8 | [139] |
| | Na | ~0.11 | 9.0 | [139] |
| | Ca | ~0.11 | 7.6 | [139] |
| 2-*Pmmn* with double vacancy | | ~0.10 | 9.2 | [139] |
| χ$_3$ | Ca | 0.23 | 7.2 | [137] |
| χ-h$_0$ | Na | 0.15 | 8.28 | [148] |
| Graphene | Li | 0.21 | 12.8 | [149] |
| Graphyne | Li | 0.19 | 13.0 | [150] |
| Silicene | Li | 0.26 | 6.35 | [151] |
| BC$_7$ | Ca | 0.26 | 4.96 | [152] |
| MoS$_2$ | Li | 0.20 | 4.80 | [153] |
| AlN | | 0.13 | 8.90 | [154] |

## 4.4 Supercapacitor

Few-layer boron sheets exhibit excellent performance as supercapacitor materials[103]. High-quality few-layer boron sheets have been synthesized in a large quantity by sonication-assisted liquid-phase exfoliation. Due to the protection of the residual solvent, the exfoliated few-layer boron sheets exhibit excellent stability away from the air oxidation. A wide potential window of up to 3.0 V and a high energy density of 46.1 Wh/Kg have been reported. After 6000 cycles, the capacitance retention of the initial specific capacitance remains about 88.7%, indicating an impressive cycling stability. This experimental study not only shows us that few layer boron sheets can be synthesized in a controlled manner but also broadens the practical application prospect of boron sheets.

## 4.5 Catalytic performances in hydrogen evolution, oxygen reduction, oxygen evolution reaction, and $CO_2$ electroreduction reaction

Borophene is the lightest catalyst for hydrogen evolution reaction (HER) [155, 156]. Wang *et al*. proposed that borophene is a high-performance catalyst for HER with near zero free energy of hydrogen adsorption, metallic conductivity and plenty of active sites in the basal plane[155]. For 2-*Pmmn* phase of borophene, due to the high surface activity, the interaction between H atom and borophene is too strong. However, for the $\chi_3$ phase of borophene, the active site is the 5-coordinated boron atoms, and as shown in Figure 9, the free energy of hydrogen adsorption $\Delta G_H$ is only 0.02 eV that is even closer to zero than that on Pt (-0.09 eV)[157]. Furthermore, the metallic band structure of $\chi_3$ phase of borophene and the multiple active sites in the basal plane are beneficial to achieve outstanding HER catalytic performance. Moreover, for $\beta_{12}$ phase of borophene, the best active site is the 4-coordinated boron atoms and the corresponding $\Delta G_H$ is only 0.1 eV. A small free energy of hydrogen adsorption $\Delta G_H$ has also been found in $\alpha$-sheet and $\beta_1$ phase of borophene. Furthermore, the high catalytic activity of borophene is not destroyed by a silver substrate. A similar conclusion has been drawn by Sun *et al*.[158] Borophene shows vast application prospect in the hydrogen evolution reaction.

Moreover, the high catalytic activity of molybdenum boride has been demonstrated experimentally.[159, 160] It is interesting that the crystal structure of $MoB_2$ contains graphene-like borophene subunits, as shown Figure 10 (a). Figure in 10 (b) shows the electrocatalytic activity of $Mo_2B$, α-MoB, β-MoB and $MoB_2$ measured in a 0.5 M $H_2SO_4$ solution at a scan rate of 1 $mV^{-1}$. The electrocatalytic activity of carbon sheet is very poor. Amorphous boron and Mo are known to have much better performances than carbon sheet. The electrocatalytic activities of the four molybdenum borides ($Mo_2B$, α-MoB, β-MoB and $MoB_2$) are even better than that of amorphous boron and Mo. The ranking of the electrocatalytic activity is amorphous B < Mo < $Mo_2B$ < α-MoB < β-MoB < $MoB_2$. The electrocatalytic activity increases with increasing boron amount. A boron-dependency of the four borides for the HER activity has been observed. Similar findings were reported recently in two phosphide systems (HER activity: Mo<$Mo_3P$<MoP)[161]. In addition, the stability measurements results of β-MoB and $MoB_2$ for the first and the 1000th cycle in 0.5m $H_2SO_4$ are shown in Figure 10 (c). The results show that $MoB_2$ and β-MoB show excellent catalytic performance in HER. The high-performance maintains well after a long-term cycle.

As shown in Figure 11, Jung *et al.* proposed that borophane supported single transition metal atoms show high catalytic performance in oxygen evolution reaction (OER) and oxygen reduction reaction (ORR).[162] The increase of the density of states near the Fermi level induced by the coupling of the *d* orbit of TM atoms and surrounding B atoms can generate active reactive sites for OER and ORR. Among the TM-BH systems, Fe–BH and Rh–BH are promising ORR electrocatalysts with overpotentials of 0.43 V and 0.47 V, respectively, whereas, for the OER, the overpotential of Rh–BH system is only 0.24 V, which is smaller than that of $RuO_2$ (0.37 V)[163]. Borophene also shows good performance as a $CO_2$ capture material[164, 165]. For the negatively charged $β_{12}$ borophene, the $CO_2$ capture capacity is up to $6.73×10^{14}$ $cm^{-2}$. Reversible $CO_2$ capture/release processes on the negatively charged $β_{12}$ phase of borophene has been observed. The charge and discharge processes can be controlled by switching on/off the charges carried by $\beta_{12}$ phase of borophene. In addition, Sun *et al*. proposed a constructive strategy to design a new catalyst for $CO_2$ electroreduction.[166] The results show that the borophene supported Cu chains possess high catalytic activity for $CO_2$ electroreduction by providing secondary adsorption sites, thus leading to small overpotentials in the preferable reaction pathway $CO_2 \rightarrow COOH^* \rightarrow CO^* \rightarrow CHO^* \rightarrow CH2O^* \rightarrow CH_3O^* \rightarrow CH3OH$.

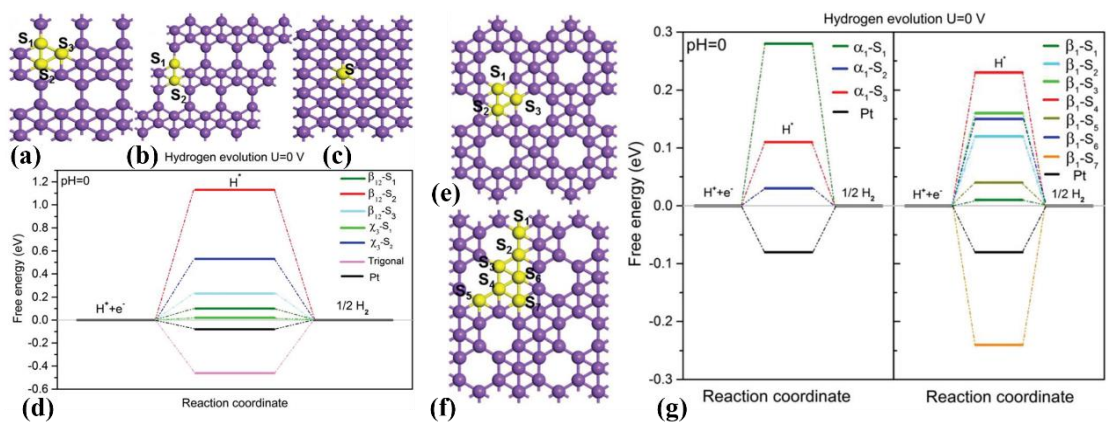

Figure 9. Crystal structures of (a) $\beta_{12}$, (b) $\chi_3$, (c) 2-*Pmmn*, (e) $\alpha$-sheet, (f) $\beta1$ borophene. (d) and (g) Free energy diagram for hydrogen evolution reactions. Reproduced from Ref.[155]

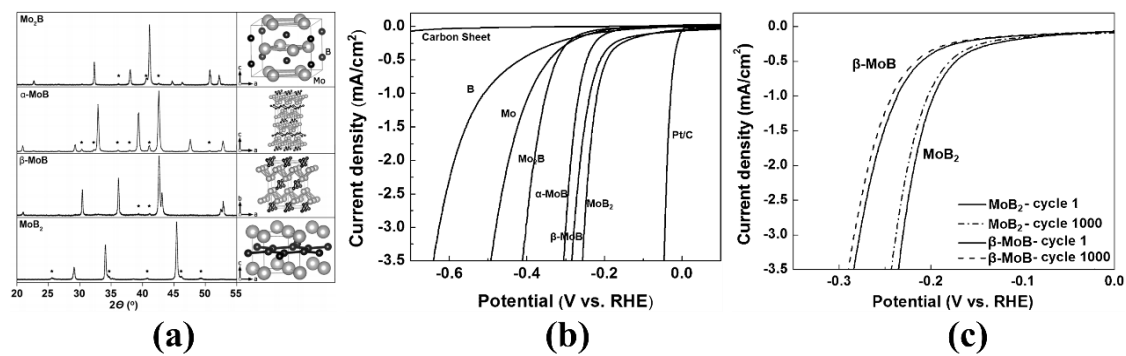

Figure 10. (a) X-ray powder diffraction patterns and crystal structures of $Mo_2B$, $\alpha$-MoB, $\beta$-MoB, and $MoB_2$. (b) Polarization curves for amorphous B, Mo, $Mo_2B$, $\alpha$-MoB, $\beta$-MoB, and $MoB_2$ measured in 0.5m $H_2SO_4$. IR-drop was corrected. (c) Stability measurements (cyclic voltammetry) of $\beta$-MoB and $MoB_2$ for the first and the 1000th cycle in 0.5 M $H_2SO_4$. The scan rate was 100 mVs1. IR-drop was corrected.[159]

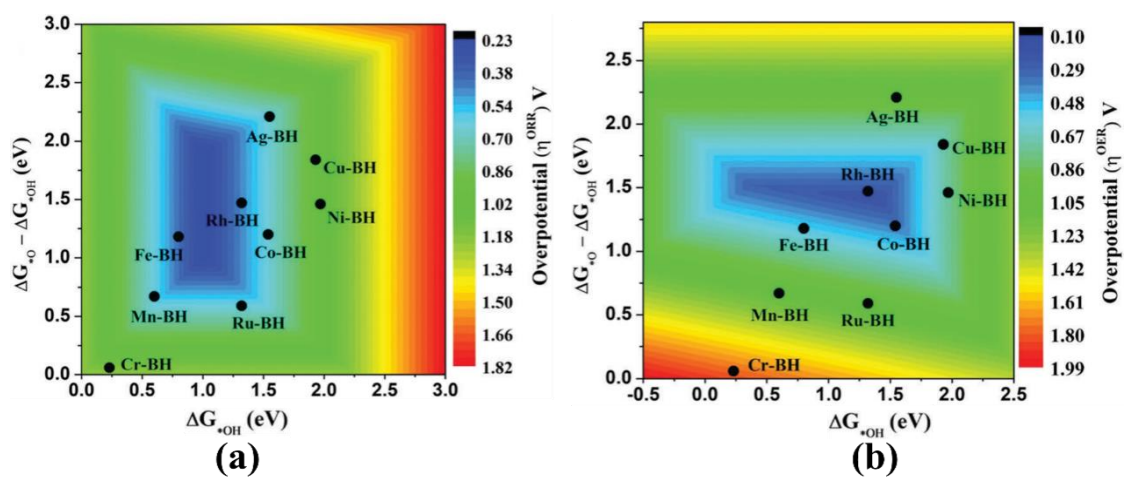

Figure 11. Thermodynamic overpotentials (in V) for (a) ORR, and (b) OER activity as a function of Gibbs binding free energies of reaction intermediates. Reproduced from Ref.[162]

## 4.6 Sensor

Borophene shows a vast application prospect in gas sensor. It has been reported that borophene can be used as a sensor for ethanol[167], formaldehyde[168] and hydrogen cyanide detection[169]. Non-equilibrium Green's function (NEGF) and density functional theory (DFT) method have been utilized to investigate the electronic properties of edge-hydrogenated 2D borophene. A sharp increase in the current passing through the borophene device is found up on adsorption of ethanol molecules. It is very similar that the electrical conductivity of $B_{36}$ borophene significantly increases at the presence of formaldehyde molecules, thereby generating an electrical signal[168]. The electronic properties of hydrogen cyanide absorbed $B_{36}$ borophene adducts are strongly dependent on the molecular adsorption configuration. Furthermore, $B_{36}$ borophene is sensitive to the concentration of HCN gas. This finding shows that borophene can be used as the sensor to detect some toxic gas. In addition, it is reported that all-boron fullerene can be used as the CO and NO sensor[170]. The electronic structure, transmission function and current-voltage of 2-*Pmmn* phase of borophene with the CO, NO, $NO_2$ or $NH_3$ adsorption has been studied by DFT and NEGF simulations[171]. The results show that the adsorption strength of CO, NO, $NO_2$ and $NH_3$ on 2-*Pmmn* phase of borophene surface is much stronger than that on graphene, $MoS_2$, and phosphorene. After the adsorption of CO, NO, $NO_2$ or $NH_3$, impressive changes of the transmission functions can be observed compared to the pristine 2-*Pmmn* phase of borophene. Furthermore, gas adsorption induced detectable quenching of I-V features, which can be regarded as OFF and ON states of the sensing mechanism, is observed.

# 5 Summary and prospect

In this review, we first introduce the experimental synthesis and theoretical prediction of borophene. The polymorphism of freestanding and the metal substrate supported borophene that is completely different from other 2D materials is particularly emphasised. A considerable number of borophene allotropes are reported whose formation energies are within a few meV/atom of the ground-state. Borophene is also completely different from other 2D materials because the distinction between borophene crystal and boron vacancy defect is blurry, due to the ultralow defect formation energy. Then, we focus on the physical and chemical properties of borophenes, mainly including the mechanical, thermal, electronic, optical and superconducting properties. We summarized the unique properties of borophene, such as the negative Poisson's ratio, highly anisotropic Young's modulus. The effects of the boron vacancy concentration, chemical modification, mechanical strain, temperature, number of layers, atomic configuration and substrate on the physical or chemical properties are presented. Finally, we review applications of borophene in the energy fields, such as the metal ion batteries, hydrogen storage, sensor and catalytic in hydrogen evolution reaction, oxygen reduction reaction, oxygen evolution reaction and $CO_2$ electroreduction reaction. Due to the light mass of boron, ultrahigh Li/Na/K/Mg/Ca/Al and hydrogen storage capacities can be achieved. Ultrafast surface ion transport along the armchair direction has been found due to the unique corrugated atomic structure of 2-*Pmmn* phase of borophene. Due to the polymorphism, outstanding catalytic performances of borophene have been found in hydrogen evolution reaction, oxygen reduction reaction, oxygen evolution reaction, and $CO_2$ electroreduction reaction. This review gives a general view of the stability, property, and application of borophenes.

Before the practical application, some key issues should be settled. Stability is one critical issue. Due to the high surface activity, borophene is prone to oxidation[51, 172]. The covalent modification is effective to improve the stability of borophene. It has been proved that hydrogenation can improve the stability of 2-*Pmmn* phase of borophene and the graphene-like borophene. Organic molecule adsorption is a feasible approach to improve air-stability. Similarly, it has been reported that the adsorption of the organic molecule on black phosphorus (BP) can improve its air-stability. The polymer-BP composites preserve both electron and hole mobility of pristine BP[173]. Furthermore, large-scale experimental growth is another issue. Now the experimental growth condition is too rigorous to large-scale growth. Suitable growth temperature, precursors and an atomically flat metal substrate need to be determined for the chemical vapor deposition, which is the first choice for the synthesis of borophene[174]. Borophene is usually synthesized on Ag/Al (111) surface. In practical application, borophene has to be transferred to the desired substrate. The adhesion energy between the borophene and substrate is weaker than that between silicene and substrate. A manner similar to mechanical exfoliation of graphene for the transfer borophene to the desired substrate has been proposed by Yakobson's research group[174].

## Acknowledgements

This work is supported by the Fundamental Research Funds for Central Universities (Grant Nos. 20720160020), Special Program for Applied Research on Super Computation of the NSFC-Guangdong Joint Fund (the second phase) under Grant No.U1501501, the National Natural Science Foundation of China (nos. 11335006, 51661135011). This work is also supported by China Scholarship Council (CSC NO. 201706310088).